\documentclass[12pt,preprint,english]{aastex}
\usepackage[T1]{fontenc}
\shorttitle{Classification of Swift's GRBs}
\shortauthors{Horvath et al.}

\begin{document}

\title{Detailed Classification of Swift's Gamma-Ray Bursts}


\author{I. Horv\'ath \altaffilmark{}}
\affil{Department of Physics, Bolyai Military University, H-1581
Budapest, POB 15, Hungary}
\email{horvath.istvan@zmne.hu}

\author{Z. Bagoly \altaffilmark{}}
\affil{Dept. of Physics of Complex Systems, E\"otv\"os University,
H-1117 Budapest, P\'azm\'any P. s. 1/A, Hungary}

\author{L. G. Bal\'azs  \altaffilmark{}}
\affil{Konkoly Observatory, H-1505 Budapest, POB 67, Hungary}
\email{}

\author{A. de Ugarte Postigo  \altaffilmark{}}
\affil{European Southern Observatory, Casilla 19001, Santiago 19, Chile \\
Osservatorio Astronomico di  
Brera (INAF-OAB), via E. Bianchi 46, I-23807, Merate (LC), Italy}
\email{}

\author{P. Veres  \altaffilmark{}}
\affil{Dept. of Physics of Complex Systems, E\"otv\"os University,
H-1117 Budapest, P\'azm\'any P. s. 1/A, Hungary \\ 
Department of Physics, Bolyai Military University, H-1581
Budapest, POB 15, Hungary}
\email{}

\and

\author{A. M\'esz\'aros \altaffilmark{}}
\affil{Faculty of Mathematics and Physics, Charles University, 
        Astronomical Institute,  V Hole\v{s}ovi\v{c}k\'ach 2, 180 00  
Prague 8,      Czech Republic }


\begin{abstract}

Earlier classification analyses found three types of gamma-ray
bursts (short, long and intermediate in duration) 
in the BATSE sample. Recent works have shown that these
three groups are also
present in the RHESSI and the BeppoSaX databases.  The duration
distribution analysis of the bursts observed by the Swift
satellite also favors the three-component model. In this paper,
we extend the analysis of the Swift data with spectral
information. We show, using the spectral hardness and
the duration simultaneously, that the maximum likelihood method
favors the three-component against the two-component model.  The
likelihood also shows that a fourth component is not needed.

\end{abstract}


\section{Introduction}

Decades ago Mazets et al. (1981) and Norris et al. (1984)
suggested that there might be a separation in the duration
distribution of gamma-ray bursts (GRBs). Kouveliotou et al.
(1993) found bimodality in the distribution of the logarithms
of the durations. Today it is widely accepted that the physics
of these two groups (short and long bursts --- called also as
Type I and Type II classes \citep{zha07,kann08,zha09,lu10})  
--- are
different, and these two kinds of GRBs are different phenomena
\citep{nor01,bal03,fox05,kann08}. The angular sky distribution
of the short BATSE's GRBs is anisotropic \citep{vav08}.
In the Swift database \citep{sak08}, the measured redshift distributions
for the two groups are also different: for short bursts the
median is 0.4 \citep{os08} and for the long ones it is 2.4
\citep{bag06}.

In the Third BATSE Catalog \citep{m6} --- using uni- and
multi-variate analyses --- \citet{ho98} and \citet{muk98} found a third
type of GRBs. Later several papers
\citep{hak,bala,rm,ho02,hak03,ho04,bor04,ho06,chat07} confirmed the
existence of this third ("intermediate" in duration) group in
the same database. The celestial distribution of this third
group in the BATSE sample is also anisotropic
\citep{me00a,me00b,li01,mgc03,vav08}.

Recent works analyzed the Swift \citep{ho08,hu09}, RHESSI
\citep{rip08,rip09} and BeppoSaX \citep{ho09} data, respectively.
They have found the intermediate class in all the three
satellites' data: in the Swift database the one-dimensional
maximum likelihood (ML) analysis of the durations has proven the
existence of these three subgroups \citep{ho08}; a preliminary
study by the same method of the BeppoSAX database \citep{fr09}
gave support for this class \citep{ho09}; in the RHESSI database
two methods led to the same results - the same one-dimensional
ML method of the durations and the bivariate ML method using
both duration and hardness \citep{rip08,rip09}.

A method to infer the physical origin of GRBs was developed
recently \citep{zha07,kann08,zha09}.
Many other observed parameters besides duration are used as
the differentiation criteria. Such a scheme only result in two
major types of GRBs (Type I and Type II). On the other hand, GRBs
may be classified using different parameters other than duration
(e.g., \citet{lu10}). We shall discuss these more in the discussion
section.

\citet{ho06} analyzed the BATSE data using duration and hardness
simultaneously; \citet{rip09} studied the RHESSI data with the
same configuration. The bivariate analysis on the Swift
database has not been done yet. \citet{ho08} only provided the
one-dimensional analysis of the durations. 

Hence, to get a
complete picture, one has to analyze the Swift data also - using
both the duration and hardness simultaneously - with the
bivariate ML method. This is the aim of this paper.

The paper is organized as follows. Section 2
briefly summarizes the method of the two-dimensional fits,
Section 3 defines the sample, Section 4 deals with these fits in
the two-dimensional parameter space and confirms the reality of
the intermediate group, Section 5 discusses the physical
differences between the classes, Section 6 contains the discussion 
 and Section 7 summarizes the
conclusions of this paper.

\section{The mathematics of the method}

When studying a GRB distribution, one can assume that the
observed probability distribution in the parameter space is a
superposition of the distributions characterizing the different
types of bursts present in the sample. Using the notations $x$
and $y$ for the variables (in a two-dimensional space) and using the law of
full probabilities \citep{renyi}, one can write
\begin{equation}\label{lfpr}
    p(x,y)
    =\sum \limits_{l=1}^k p(x,y|l)p_l.
\end{equation}
In this equation, $p(x,y|l)$ is the conditional probability
density assuming that a burst belongs to the $l$th class. $p_l$
is the probability for this class in the observed sample ($\sum
\limits_{l=1}^k p_l = 1$), where $k$ is the number of classes.
In order to decompose the observed probability distribution
$p(x,y)$ into the superposition of different classes, we need the
functional form of $p(x,y|l)$. The probability distribution of
the logarithm of durations can be well fitted by Gaussian
distributions, if we restrict ourselves to the short and long
GRBs \citep{ho98}. We assume the same also for the $y$
coordinate. With this assumption we obtain, for a certain $l$-th
class of GRBs,
$$
p(x,y|l)  =
 \frac{1}{2 \pi \sigma_x \sigma_y
\sqrt{1-r^2}} \times \;\;\;\;\;\;\;\;\;\;\;\;\;\;\;\;\;\;\;\;
\;\;\;\;\;\;\;\;\;\;\;\;$$
\begin{equation} \label{gauss}
\exp\left[-\frac{1}{2(1-r^2)}
\left(\frac{(x-a_x)^2}{\sigma_x^2} + \frac{(y-a_y)^2}{\sigma_y^2}
- \frac{2r(x-a_x)(y-a_y)} {\sigma_x \sigma_y}\right)\right], \;
\end{equation}
where  $a_x$, $a_y$ are the means, $\sigma_x$, $\sigma_y$ are
the dispersions, and $r$ is the correlation coefficient
\citep{tw53}. Hence, a certain class is defined by five independent
parameters, $a_x$, $a_y$, $\sigma_x$, $\sigma_y$, and $r$, which are
different for different $l$. If we have $k$ classes, then we
have $(6k - 1)$ independent parameters (constants), because any
class is given by the five parameters of Eq.(\ref{gauss}) and
the weight $p_l$ of the class. One weight is not independent,
because $\sum \limits_{l=1}^{k} p_l = 1$. The sum of $k$
functions defined by Eq.(\ref{gauss}) gives the theoretical
function of the fit.

\begin{figure}
 \includegraphics[width=12.1cm, angle=0]{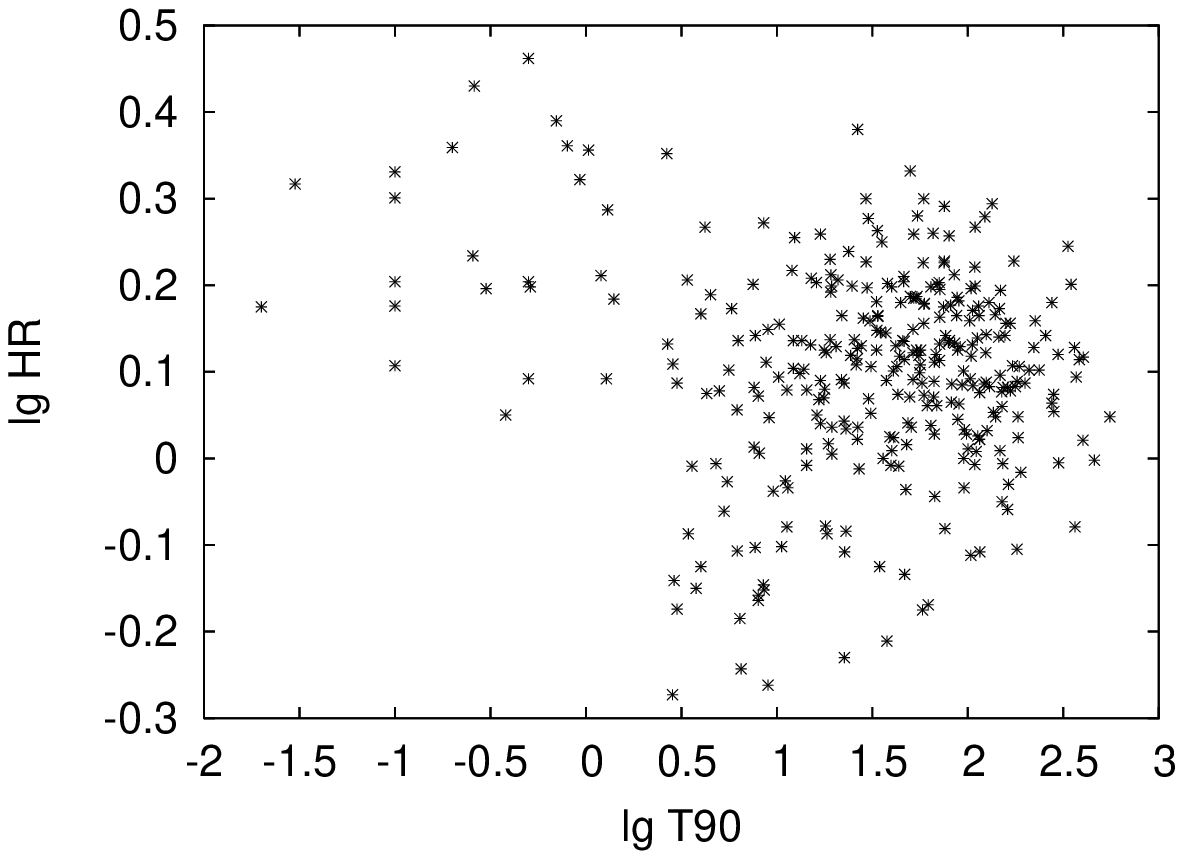}\\
  \caption{Distribution of $N = 325$  GRBs in the 
$\{\lg T_{90}; \lg HR\}$ plane. } \label{t90h32}
\end{figure}

By decomposing $p(x,y)$ into the superposition of $p(x,y|l)$
conditional probabilities one divides the original population of
GRBs into $k$ groups.
Decomposing the left-hand side of Eq.(\ref{lfpr}) into the sum
of the right-hand side, one needs the functional form of
$p(x,y|l)$ distributions, and also $k$ has to be fixed. Because
we assume that the functional form is a bivariate Gaussian
distribution (see Eq.(\ref{gauss})), our task is reduced to
evaluating its parameters, $k$ and $p_l$.

\citet{bal03} used this method for $k=2$, and gave a more
detailed description of the procedure. 
 However, that paper used fluence instead of
hardness, and used the BATSE data. Also \citet{ho06} used this
method analyzing the BATSE data. Here we will make similar
calculations for $k=2$, $k=3$ and $k=4$ using Swift
observations.

\section{The sample} 

In the Swift BAT Catalog \citep{sak08} there are 237 GRBs, of
which 222 have duration information. Following the 
same procedure of data reduction we have extended this
sample with all the bursts detected until mid 2008 December 
(ending with GRB 081211a). 
Our total sample thus comprises the first four years of the
Swift satellite (since the detection of its first burst GRB
041217) and includes $325$ bursts. $222$ from \citet{sak08} and
$103$ reduced by us. The data reduction was done by using
HEAsoft v.6.3.2 and calibration database v.20070924. For
light curves and spectra we ran the \texttt{batgrbproduct}
pipeline 
\footnote{ http://heasarc.nasa.gov/lheasoft/ }
We fitted the spectra integrated for
the duration of the burst with a power-law model and a power-law
model with an exponential cutoff. As in \citet{sak08} we have
chosen the cutoff power-law model if the $\chi^2 $ of the fit
improved by more than 6.

For calculating the hardness ratio, we have chosen fluence 2
($25-50 keV$) and fluence 3 ($50-100 keV$) and the hardness is
defined by the $HR = F3 / F2 $ ratio.
Figure \ref{t90h32} shows the distribution of GRBs in the
$\{\lg T_{90}; \lg HR\}$ plane, where the fits were made 
for $x = \lg T_{90}$ and $y = \lg HR$.

\section{Bivariate ML fitting} 

In order to find the unknown constants in Eq.(\ref{gauss}), we
use the maximum likelihood (ML) procedure of parameter
estimation \citep{bal03}. Assuming a set of $N$ observed $[x_i,
y_i], \, (i=1, \dots, N)$ values ($N$ is the number of GRBs in
the sample for our case, which here is 325) we can define the
likelihood function in the usual way, after fixing the value of
$k$, in the form 
\begin{equation}\label{ml}
  L=\sum \limits_{i=1}^N \ln p(x_i,y_i)\;,
\end{equation}
where $p(x_i,y_i)$ has the form given by Eq.(\ref{lfpr}).
Similarly to what was done by \citet{{bal03}} and \citet{{ho06}},
the EM (Expectation and Maximization) algorithm is used to
obtain the $a_x, a_y, \sigma_x, \sigma_y, r$, and $p_l$
parameters at which $L$ reaches its maximum value. 
We made the calculations for different values of $k$ in order to
see the improvement of $L$, as we increase the number of
parameters to be estimated. Tables \ref{tab1}-\ref{tab2}
summarize the results of the fits for $k=2$ and $k=3$.
Figures 2 and 3 show the results in 
the $\{\lg T_{90}; \lg HR\}$ plane.

\begin{figure}
 \includegraphics[width=12.1cm, angle=0]{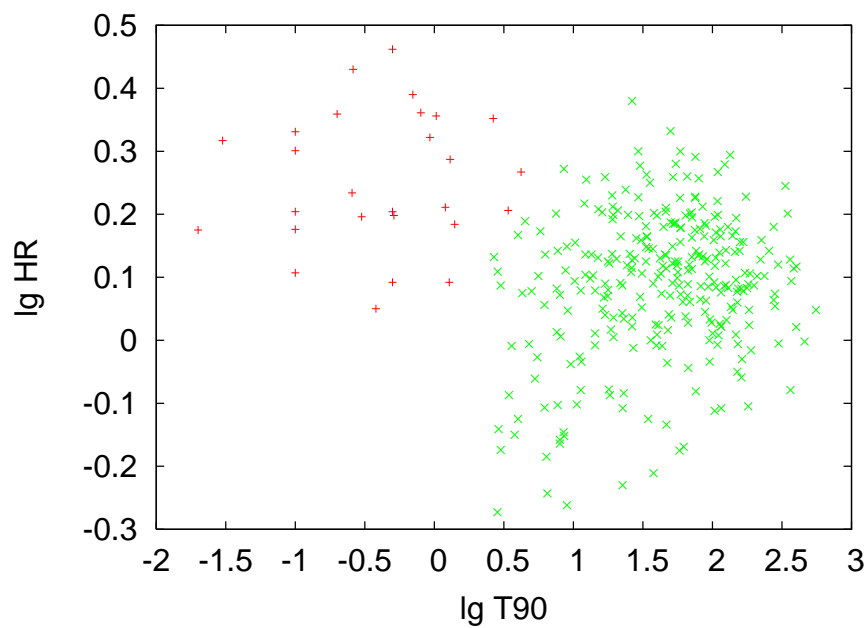}\\
  \caption{Distribution of $N = 325$  GRBs in the 
$\{\lg T_{90}; \lg HR\}$ plane. The different symbols mark bursts
belonging to the short (red +) and long (green x) classes,
respectively.} \label{k2}
\end{figure}

\begin{table}  \centering  \caption{Results of the EM algorithm in the 
$\{\lg T_{90}; \lg HR\}$ plane.
  $k=2$ \ \ }
\label{tab1}
  \begin{tabular}{ccccccc} \hline
   $l $ & $p_l$  &  $a_x $ &  $a_y$ & $\sigma_x$ & $\sigma_y$  &  $r$ \\
 \hline
 1  &  0.082 &  -0.383  &  0.256  &  0.602  &  0.114  &  0.071  \\
  2 & 0.918  &  1.628  &  0.096  &  0.516  &  0.117  &  0.226 \\
 \hline
\end{tabular}
\end{table}

The confidence interval of the estimated parameters  can be
given on the basis of the following theorem. Denoting by
$L_{max}$ and $L_0$ the values of the likelihood function at the
maximum and at the true value of the parameters, respectively,
one can write asymptotically as the sample size
$N\rightarrow\infty$ \citep{Kendall76},
\begin{equation}\label{lmax}
    2(L_{max}-L_0)\approx \chi_m^2
\end{equation}
where $m$ is the number of estimated parameters ($m=6k -1$ in
our case), and $\chi_m^2$ is the usual $m$-dimensional $\chi^2$
function \citep{tw53}. Moving from $k=2$ to $k=3$ the number of
parameters $m$ increases by 6 (from 11 to 17) and $L_{max}$
grows from 506.6 to 531.4. 
Since $\chi_{17}^2=\chi_{11}^2+\chi_6^2$ the increase in
$L_{max}$ by a value of 25 corresponds to a value of 50 for a
$\chi_6^2$ distribution. The probability for $\chi_6^2 \geq 50$
is very low ($10^{-8}$), so we may conclude that the inclusion
of a third class into the fitting procedure is well justified by
a very high level of significance.

Moving from $k=3$ to $k=4$, however, the improvement in
$L_{max}$ is 3.4 (from 531.4 to 534.8) corresponding to
$\chi_6^2 \geq 6.8$, which can happen by chance with a
probability of 33.9 \%. 
Hence, the inclusion of the fourth class is {\em not} justified.
We may conclude from this analysis that the superposition of
three Gaussian bivariate distributions - and {\em only these
three ones} - can describe the observed distribution.

\begin{figure}
 \includegraphics[width=12.1cm]{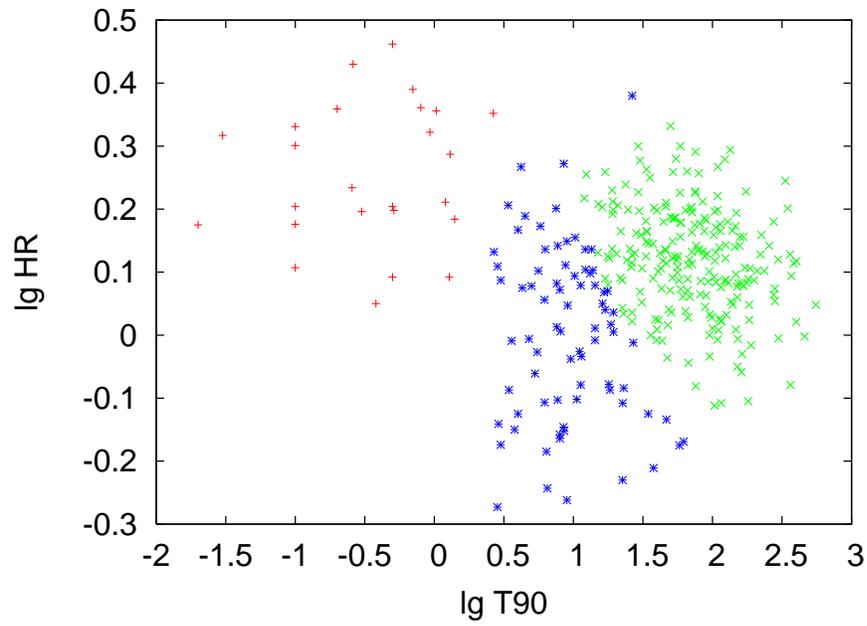}\\
  \caption{Distribution of $N = 325$ GRBs in the $\{\lg T_{90}; 
\lg HR\}$ plane. The different symbols mark bursts
belonging to the short (red +), intermediate (blue *) and long (green x) classes,
respectively.} \label{k3}
\end{figure}

\begin{table}
  \centering
  \caption{Results of the EM algorithm.  $k=3$ \ \   }\label{tab2}
  \begin{tabular}{ccccccc} \hline
  $l $ & $p_l$  &  $a_x $ &  $a_y$ & $\sigma_x$ & $\sigma_y$  &  $r$ \\
\hline
 1 & 0.079 &  -0.426  &  0.259  &  0.576  &  0.114  &  0.120 \\
 2 & 0.296 &  1.076 & 0.025  &  0.376  &  0.129 &  -0.004  \\
 3  &  0.626 &  1.882  &  0.130  &  0.350  &  0.093  & -0.237  \\
   \hline
\end{tabular}
\end{table}

This means that the 17 independent constants for $k=3$ in Table
\ref{tab2} define the parameters of the three groups. We see
that the mean hardness of the intermediate class is very low -
the third class is the softest one. This is in a good agreement
with \citet{ho06}, who found that the intermediate duration class
is the softest in the BATSE database. In that database,  
11\% of all GRBs belonged to this group. 
In our analysis, $p_2 =0.296$; therefore, 30\% 
of the Swift bursts belong to the third group.

\section{Separation of GRBs into the classes}

Based on the calculations in the previous paragraph, we resolved
the $p(x,y)$ probability density of the observed quantities into
a superposition of three Gaussian distributions. Using this
decomposition, we can classify {\it any} observed GRB into the
classes represented by these groups (this is similar to the
\citet{ho06} work dealing with the BATSE data). In other words,
we develop a method allowing us to obtain, for any given GRB,
its three membership probabilities, which define the likelihood
of the GRB to belong to the short, intermediate, and long
groups. The sum of these three probabilities is unity. For this
purpose we define the following $I_l(x,y)$ indicator function,
which assigns to each observed burst a membership probability in
a given $l$ class as follows:
\begin{equation}\label{mprob}
    I_l(x,y)= \frac{ p(x,y|l) p_l}{\sum \limits_{l=1}^k  p(x,y|l) p_l}.
\end{equation}

According to Eq.(\ref{mprob}), each burst may belong to any of
the classes with a certain probability. In this sense, one cannot
assign a given burst to a given class with absolute certainty,
but with a given probability. This type of classification is
called a "fuzzy" classification \citep{mclaba88}. Although, any
burst with a given $[x,y]$ could be assigned to all classes with
a certain probability, one can select that $l$ at which the
$I_l(x,y)$ indicator function reaches its maximum value. 
For $k=3$, Figure 3 shows the bursts' distribution in
the $\{\lg T_{90}; \lg HR\}$ plane indicating the group memberships.
For the community, we make the list of the membership probabilities 
available on the internet.
\footnote{ The list of the membership probabilities can be found at 
http://itl7.elte.hu/$\sim$veresp/sak325T90H32CL3.res}
Table \ref{prob} contains the six GRBs identified as intermediate
having redshift information.

\begin{table}
  \centering
  \caption{ The six intermediate GRBs identified with redshift.
\ \   }\label{prob}
  \begin{tabular}{ccccccc} \hline
 GRB id. & 3rd type prob.   & redshift \\
\hline
050525 & 0.78 &  0.606  \\
050922C & 0.71 & 2.17   \\
060206 &  0.88 &  4.048    \\
071117 &  0.82 &  1.331    \\
080913 &  0.67 &  6.695    \\
081007 &  1.00 &  0.5295    \\
 \hline
\end{tabular}
\end{table}

One can demonstrate the robustness of classification by
comparing the results obtained from the $\{\lg T_{90};
\lg HR\}$ and the $\{\lg T_{50};
\lg HR\}$ planes. A cross tabulation
between these two classifications is given in Table \ref{tabcont}.
According to the table, the short and long
classes correspond within a few percent to the respective groups
obtained from the other classification. We even have three zeros,
which means, for example, no short burst, classified with $\{\lg T_{90}; \lg HR\}$,
was identified as long with the $\{\lg T_{50}; \lg HR\}$ 
classification and since the other number is also zero, this applies vice versa.
Consequently, the
robustness of the short and long groups is well established. On the
other hand, the population of the intermediate group is less numerous
when classifying in the $\{\lg T_{50}; \lg HR\}$ plane
than in the other one. Table \ref{tabcont} clearly shows that
the $\{\lg T_{90}; \lg HR\}$ classification, except for three
(two for long and one for short), contains all
GRBs assigned to the intermediate group by $\{\lg T_{50}; \lg HR\}$.

\begin{table}
  \centering
  \caption{ The cross-tabulation of the two classifications (with  
$T_{90}-HR$ and $T_{50}-HR$).   \ \   }\label{tabcont}
  \begin{tabular}{ccccccc} \hline
 T50 $\backslash$ T90 & short ($T_{90}-HR$) & interm. ($T_{90}-HR$) & long ($T_{90}-HR$) & total \\
\hline
 short ($T_{50}-HR$)  & 24 &  0 &  0 & 24 \\
interm. ($T_{50}-HR$) & 1 &  62 & 2 & 65 \\
 long ($T_{50}-HR$)   &  0 &  24  &  212  & 236 \\
\hline
total &  25 &  86  &  214  & 325 \\
 \hline
\end{tabular}
\end{table}

The intermediate bursts
classified in the $\{\lg T_{90}; \lg HR\}$
plane indicated 0  GRBs from the short and 24 from the long
group identified in the other plane. This moderately high number of indicated bursts clearly
shows that a slight variation of the parameters of the Gaussian
distribution representing the intermediate group results in a
moderate change in the number of identified objects in this group.
Comparing the  number of GRBs belonging to the
intermediate group, one gets 86 and 65; 
86 are identified as intermediate with the $T_{90}$
analysis and 65 are identified as intermediate with the $T_{50}$ 
analysis (see the numbers in Table \ref{tabcont}).

If one assigned the burst to that group that had the maximum
membership probability, a slight change in the parameters of the
corresponding Gaussian distribution may move the GRB to an other group. On the
contrary, the fuzzy classification assigns membership probability
to all of the bursts. Hence, a small variation of the parameter gives a
small variation in the estimated number of bursts in the
intermediate group obtained by summing the membership
probabilities of all GRBs in the sample.
There is a further issue to be considered here. The duration
($T_{90}$) is   dependent on the energy range at which we are
measuring. This results in a further fuzziness factor in the determination of the
 membership.

\section{Discussion }

\subsection{Redshift Distributions}

\begin{figure}
 \includegraphics[width=9.1cm, angle=-90]{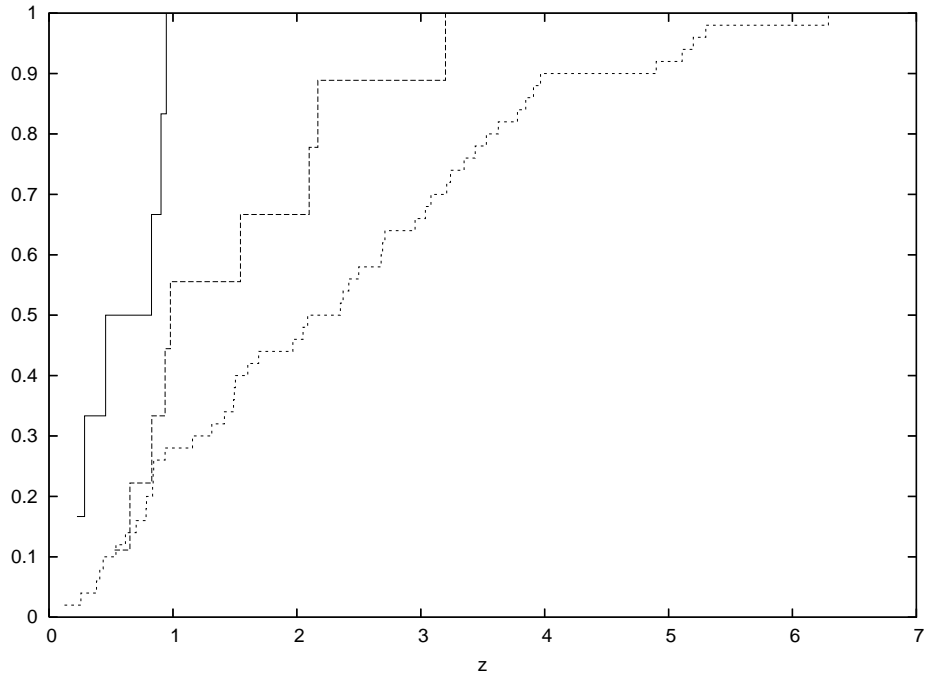}\\
  \caption{Cumulative redshift distribution of the three classes. 
The continuous line is the short, the dashed line is the intermediate,
and the dotted line is the long population.} \label{z1}
 \end{figure}

The cumulative redshift distribution of the three populations is shown in Figure \ref{z1}.
Redshifts were taken from \cite{ama08}. Only a subset of the classified bursts
had redshift information and we considered bursts where the probability of
belonging to a given population is higher than $97\%$. This means $6$ short,
$9$ intermediate, and $50$ long GRBs. The long and short population
redshift distributions are significantly different ($99.4\%$ significance). 
The intermediate GRBs redshift distribution is clearly between the
short and long redshift distributions, which could mean that they are
further than the short bursts and closer than the long ones.
However, probably owing to the small number of data points the
difference is not significant. We have tried several statistical tests, such as  
Kolmogorov-Smirnov, Wilcoxon, and Mann-Whitney.
None of them showed high significance; the best one
was $92\%$.
Therefore, we are not able to prove that the long and the
intermediate bursts came from different redshift distributions.
A more 
extensive
analysis on these aspects as well as the afterglow properties of the 
intermediate
group of events will be presented in a parallel work \citep{adeu10}. 

\subsection{The 
$E_{\mathrm{peak}}-E_{\mathrm{iso}}$ (Amati) Relation }

Once we classify the bursts, it is also possible to investigate their properties in
the context of the $E_{\mathrm{peak}}-E_{\mathrm{iso}}$ or Amati-relation
\citep{ama08} in the case of bursts with measured redshift. The Amati-relation \citep{ama02} 
is a correlation between the rest-frame peak energy of the GRB spectrum
($E_{\mathrm{peak}}$) and the isotropic-equivalent energy release of the
burst($E_{\mathrm{iso}}$). 

Again, only a fraction of the populations had redshift and could be placed on
the $E_{\mathrm{peak}}-E_{\mathrm{iso}}$ plane. We have $18$ from the long
population and $6$ from the intermediate. 
Both groups seem to
follow the same relationship. As the
Amati-relation is not valid for the short population, the intermediate
bursts are more closely related to the long population than to the short class. 

\begin{figure}
 \includegraphics[width=12.1cm, angle=0]{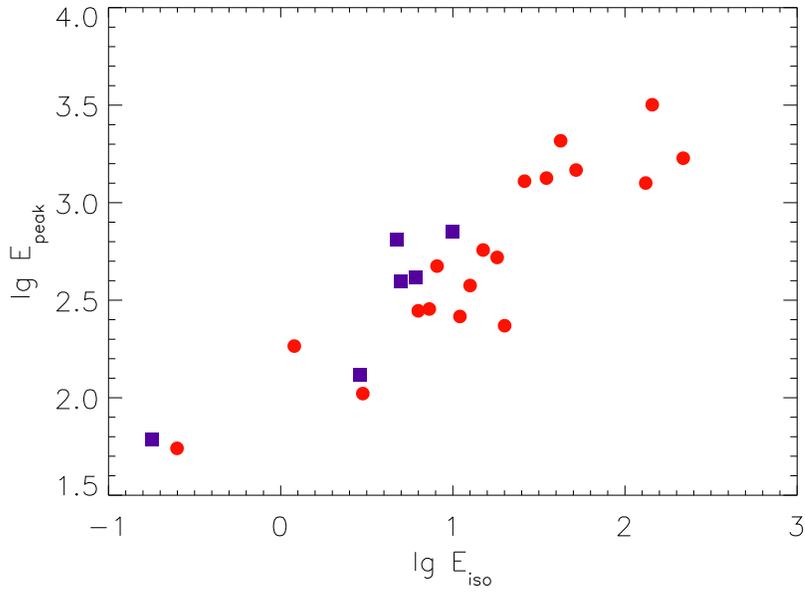}\\
\caption{$E_{\mathrm{peak}}-E_{\mathrm{iso}}$ relation of the classified
bursts. Red symbols represent long GRBs, and intermediate bursts are
plotted in blue.
The y-axis is the 10-base logarithm of $E_{\mathrm{peak}}$ in keV, and the
x-axis is the 10-base logarithm of $E_{\mathrm{iso}}$ in units of
$10^{52}$ erg/s. } \label{am1}
 \end{figure}

Intermediate bursts do not populate the most energetic regime of the
$E_{\mathrm{peak}}-E_{\mathrm{iso}}$ plane unlike the long bursts (see Figure
\ref{am1}). They tend to have lower isotropic energies compared to the long
population. The small number of data points makes hard to give firm assertions at
this time. Also, there is no significant clustering of intermediate bursts on
this plane. 

\subsection{Relation to Other Classifications}

One can compare this classification to the Type I/II method put forward by
\cite{zha07} and \cite{zha09}. The Type I/II scheme looks for signatures of the binary merger
and the collapsar scenario (as association with supernova, host galaxy
properties, spectral lag etc.) in bursts and classifies them accordingly.
Currently these two scenarios are thought to be the most probable progenitors
of GRBs. This method uses a wide range of observations for classification (some
deemed more decisive than others) and as such only a fraction of the bursts can
be assigned Type I or II.
 \cite{zha09} publish a table with the most certain members
called the Gold sample. Though this scheme allows only for two classes
it is worth checking their membership in our classification scheme.

There are five bursts of Type I in the Gold sample. GRB050709 was detected by 
HETE II so it is not included in our sample.
The other four GRBs are listed in Table \ref{tabx}.
This table contains the group membership probabilities of the bursts from the Type I Gold sample 
of \cite{zha09}, which clearly shows none of them belong to the intermediate class.

\begin{table}
 \centering
 \caption{Membership probabilities of the Type I Gold sample of
\cite{zha09} (columns 2-4). The memberships calculated
without the extended emission (EE) are in the last three columns
(marked with a *). }\label{tabx}
 \begin{tabular}{ccccccc} \hline
 GRB name & P(short)  & P(interm.)  & P(long) & P$^*$(short) &
P$^*$(interm) & P$^*$(long)    \\
\hline
 050509B & 1  &  0   &  0 & 1 & 0 & 0\\
 050724  & 0  &   0.02 &   0.98 & 0.02 & 0.95 & 0.03\\
 060614  & 0  &   0.02 &   0.98 & 0.01 & 0.98 & 0.01\\
 061006  & 0  &   0.01 &   0.99 & 0.53 & 0.44 & 0.03\\
\hline
\end{tabular}
\end{table}

The lightcurves of the last three of the bursts in Table \ref{tabx} have a
short-hard  spike followed by an extended emission tail. If we disregard the
extended emission we get a different duration. The intermediate group
membership probability increases in this case to $\sim 0.95$, $0.98$ and $\sim 0.45$ for
050724, 060614 and 061006 respectively. 061006 has a $\sim 0.55$ probability of
belonging to the short group. Care must be taken with these probabilities as
only a whole new classification with the new durations would yield the correct
membership probabilities.

GRB 060218 and  X-ray flash 080109 have associated supernovae. Unfortunately, there is no
duration available for any of them. For 060218, a minimum duration of $\sim2700
$ s can be established \citep{g060218}, and the logarithm of hardness ratio of
$\lg HR=-0.15$ can be derived. These values place this burst to the extreme
right of the hardness-duration diagram and thus it belongs to the long
population (though it is very soft). 080109 was detected with XRT and BAT
only measured upper limits; therefore, we cannot infer anything about its group
membership.

\section{ Conclusion}

In the BATSE analysis \citet{ho06} found the intermediate type of
burst to be the softest. 
To be sure that the 
classifications are free of instrumental
effects, it is important to compare the results obtained with datasets 
from different satellites. This is one of the
main aims of this paper.
We find that the bursts
observed by Swift can be divided (in the duration hardness plane)
by three groups and only three
groups as happened with the BATSE sample.

Our results, summarized in Table 2, are very similar to
\citet{ho06} results obtained using the BATSE data. 
This indicates that 
the two satellites are observing a similar population of bursts.
The relative size of the different groups will differ with the
detector parameters: the Swift observations are more sensitive
for soft and weak bursts; hence, the observation probability of
the intermediate group members is slightly enhanced compared to
BATSE (from 11\% to 30\% ).

An important question that must be answered in this context is
whether the intermediate group of GRBs, obtained in the previous
paragraph from the mathematical phenomenological classification, really represents
a third type of burst physically different from both the short
and the long population.

To infer the physical origin of a GRB, one must collect more
direct information about the GRB progenitors. For example, most
long GRBs are found to have irregular host galaxies with intense
star formation \citep{fru06} and some are associated with
a supernova (\citet{pian06}, and references therein). Some short
GRBs are associated with nearby galaxies with low star formation
rate \citep{geh05,fox05,ber05,zha07}, 
which point toward a possible origin of
compact star mergers.

Since one has only two main types of suggested progenitors,
massive star progenitors for long (Type II) and compact star
mergers for short (Type I), there are two possibilities:

A, There are only
two physically different types of GRBs, and the statistically significant
third group belongs to one of them.

B, The third group is physically real, and one
should look more carefully at the new observations
to find this new type of progenitor.

According to this paper's analysis, the intermediate GRBs are the softest
among the three classes. This different small mean hardness and
also the different average duration suggest that the
intermediate group should also be a different phenomenon, that
is, both in hardness and in duration the third group differs
from the other two. 

To sum up, our bivariate ML method confirmed the existence
of the third intermediate subgroup on the high significance
level in the Swift database. Existence of other groups is not supported.
It is conjectured that the GRBs of intermediate subgroup are physically
different phenomena.

\acknowledgments

This research was supported in part through 
OTKA K077795 grant, by the GAUK grant No. 46307, 
by the GA\v{C}R grant No. P209/10/0734, 
by OTKA/NKTH A08-77719 and A08-77815 (ZB),
by the Research Program MSM0021620860 of the Ministry
of Education of the Czech Republic (AM), 
by ASI grant SWIFT I/011/07/0, by the Ministry of University 
and Research of Italy (PRIN MIUR 2007TNYZXL) (AdUP),
by a Bolyai Scholarship (IH), and
by an ESO fellowship (AdUP).
 Thanks are due to the valuable remarks to Jakub \v{R}\'{\i}pa.
The detailed and constructive suggestions and remarks of the
anonymous referee significantly helped to improve the scientific merit
of this paper.

\clearpage

\end{document}